\pdfoutput=1

\documentclass[ ]{aa}
\usepackage{ifpdf}
\usepackage{txfonts}
\usepackage{graphicx}

\usepackage[tight,scriptsize]{subfigure}

\usepackage{float}

\usepackage{enumitem}

\usepackage{natbib}

\newcommand{\metss}{\,\mathrm{m\,s^{-2}}}

\newcommand{\kms}{km~s$^{-1}$}
\newcommand{\cc}{{cm$^{-3}$}}
\newcommand{\DNS}{DN~s$^{-1}$}
\newcommand{\cmsq}{{cm$^{-2}$}}

\newcommand{\HI}{\ion{H}{i}}
\newcommand{\HeI}{\ion{He}{i}}
\newcommand{\HeII}{\ion{He}{ii}}

\newcommand{\CIV}{\ion{C}{iv}}

\newcommand{\FeVIII}{\ion{Fe}{viii}}
\newcommand{\FeIX}{\ion{Fe}{ix}}
\newcommand{\FeXII}{\ion{Fe}{xii}}
\newcommand{\FeXIV}{\ion{Fe}{xiv}}

\begin{document}

\title{Analysis of UV and EUV emission from impacts on the Sun after 2011 June 7 eruptive flare}

\author{D.E. Innes\inst{1,2}
\and P. Heinrich\inst{1,3}
\and B. Inhester\inst{1}
\and L.-J. Guo\inst{1,2}}
\institute{Max-Planck-Institut f\"{u}r Sonnensystemforschung, 37077 G\"ottingen, Germany
\and Max Planck/Princeton Center for Plasma Physics, Princeton, NJ 08540, USA 
\and Institute of Physics, Carl von Ossietzky University Oldenburg, 26129 Oldenburg, Germany}

\abstract
{On 2011 June 7 debris from a large filament  eruption fell back to the Sun causing bright ultraviolet (UV) and extreme ultraviolet (EUV) splashes across the  surface. These impacts may give clues on the process of stellar accretion. }
{The aim is to investigate how the impact emission is influenced by structures in the falling ejecta and at the solar surface.}
{We determine the UV and EUV light curves of a sample of impacts.  The ballistic impact velocity is estimated from the ejection and landing times and, where possible, compared with the velocity derived  by tracking the downflows in SDO/AIA and STEREO/EUVI images. Estimates of the column density before impact are made from the darkness of the falling plasma in the 193~\AA\ channel. }
{The impact velocities were between 230 and 450~\kms. All impacts produced bright  EUV emission at the impact site but bright UV was only observed when the impacting fragments reached the chromosphere. There was no clear relation between EUV intensity and kinetic energy.  Low UV to EUV intensity ratios (I$_{UV}$/I$_{EUV}$) were seen (i) from impacts of low column-density fragments, (ii) when splashes, produced by some impacts, prevented subsequent fragments from reaching the chromosphere, and (iii)  from an impact in an active region.The earliest impacts with the  lowest velocity ($\sim250$~\kms) had the highest I$_{UV}$/I$_{EUV}$. }
{
The I$_{UV}$/I$_{EUV}$ decreases with impact velocity, magnetic field at the impact site, and EUV ionising flux. Many of the infalling fragments
dissipate above the chromosphere either due to ionisation and trapping in  magnetic structures, or to them encountering a splash from an earlier impact. 
If the same happens in accreting stars then the reduced X-ray compared to optical emission that has been observed is more likely due to absorption by the 
trailing stream than locally at the impact site.
}


\titlerunning{Emission from splashes onto the Sun}


\maketitle
\section{Introduction}
The  plasma falling and crashing into the Sun after the eruptive flare on 2011 June 7 has been compared to stellar accretion flows \citep{Reale13}. The velocities and densities of impacting fragments are similar. Also in both situations the magnetic field may affect the flows as they fall towards the surface. The impacts during the 2011 June 7 event fell with a range of speeds and densities into a variety of magnetic environments.  
They may therefore provide a testbed for accretion onto unresolved stellar objects. 


Accretion onto young stars is thought to explain  excesses seen in their infrared \citep{Mendoza66,Rucinski85}, ultraviolet (UV) \citep{Herbig86} and soft X-ray \citep{Kastner02,Schmitt05} emission compared to stars of similar spectral type. 
Models of the emission from accretion have made a number of critical assumptions regarding the structure of the accretion flows, the pre-impact sites, and the optical thickness of the heated plasma at the impact site. 
For example, simulations often assume uniform plasma columns channelled along one-dimensional (1D) flux tubes \citep{Calvet98, Sacco08, Sacco10} or accretion  along uniform vertical fields \citep{Orlando10, Bonito14}. 
 \citet{Reale13} used the spatially resolved solar impacts on June 7 to investigate the applicability of 2D hydrodynamic (HD) accretion models to prominence plasma falling back to the Sun. They investigated fragments with a velocity around 400~\kms\ and a density $10^{10}$~\cc\ plunging through  the solar corona and into chromosphere.
 Similar to the stellar accretion models, the impacting fragments sank deep into the chromosphere where their energy was dissipated creating forward and reverse shocks with high-temperature outflows from the impact region into the surrounding corona. 
  To model the emission 
  \citet{Reale13} assumed all EUV emission from below the transition region was absorbed and therefore not detected. With this assumption, they concluded that the
observed EUV intensities originated from the 5 to 30\% of the fragment plasma that radiated above the transition region.  
A study of stellar accretion revealed that accretion rates derived from X-ray fluxes are considerably lower than those inferred from UV/optical/ infrared fluxes \citep{Curran11}, suggesting that 
in stellar systems X-rays may also be absorbed. \citet{Reale13} argue that  absorption at the impact site may  explain the lower accretion rates derived from X-ray observations; however
their models did not account for structure induced by magnetic fields at the impact site.   Magnetic fields may reduce the  
 local absorption \citep{Orlando13,Bonito14}, leading \citet{Bonito14} to conclude that the unperturbed, trailing accretion stream rather than absorption at the impact site is a more probable explanation for the low X-ray fluxes. 

The solar observations of the June 7 impacts can be used to investigate impact depth and effects of magnetic structuring at the impact site by comparing high resolution EUV  (131, 171, 193, 211~\AA)  and UV (1600 and 1700~\AA)  observations of the impacts because the  EUV and UV emission come from different depths in the atmosphere. The EUV emission is only visible above the transition region because EUV from below is obscured by \HI, \HeI, \HeII. On the other hand both the transition region and chromosphere are transparent to UV emission.  Thus the UV intensities give a indication of how much plasma penetrates to the chromosphere. Naively one would expect that since the EUV from deep impacts is absorbed, then the intensity of the UV to EUV (I$_{UV}$/I$_{EUV}$) should increase the deeper the impact. Also according to \citet{Sacco10},  the faster impacts should penetrate deeper into the atmosphere. Consequently one expects the (I$_{UV}$/I$_{EUV}$) to increase with velocity. 
By surveying the UV and extreme ultraviolet (EUV) emission from solar impacts of different sizes and velocities, we investigate the conditions controlling the optically thin (UV) to optically thick (EUV) intensity ratios at impact sites on the Sun. 

The eruption on 2011 June 7 in NOAA AR 11226 (S22W55) resulted in a fast coronal mass ejection (CME) with velocity about 1250~\kms\  in interplanetary space. In EUV images,  a dome-shaped CME wave expanded  across the disk with velocities of up to 960~\kms\ \citep{Li12, Cheng12}. Behind the fast wave, filament plasma was thrown upward, and a lot of it fell back creating impacts across the solar surface.  Four of the impacts into the quiet Sun have been studied by \citet{Gilbert13}. In each of these, the main brightening is due  to dissipation of kinetic energy on impact. The eruption was seen from both the Solar Dynamics Observatory (SDO) and the Solar TErrestial Relations Observatory Ahead (STEREO-A) spacecraft when STEREO-A was 94$^\circ$ ahead of Earth.  \citet{Gilbert13} used images from both these spacecraft to estimate the height-time progress of several ejecta. Although straightforward  in theory, in practice it is difficult  to pick out exactly the same  ejecta from both viewpoints in the image pairs with 2.5~min cadence, especially when the material is close to the Sun. The velocities given by \citet{Gilbert13} were determined roughly 5~min before impact  which means that the actual impact velocity is probably about 50-100~\kms\ higher, given the solar gravitational acceleration close to the surface, $274 \metss$. 
   
The coronal magnetic field was complex with connections between the three neighbouring active regions in the south \citep{vandriel14} and the active region to the north. As the plasma falls back through the corona, it breaks up due to Rayleigh-Taylor type of instabilities \citep{Innes12} generating long, thin strands with dense clumps at their heads \citep{Carlyle14}.

In this paper we present the results of a survey of the UV and EUV emission from 16 fragments. In the next section, we describe the  image sequences used. The analysis first explains the basis for the impact velocity, then in the next section we present the characteristics of the 16 impacts, and show details of fragments impacting into various environments. The discussion section summarises the results and in particular the conditions for large UV to EUV intensity ratios (I$_{UV}$/I$_{EUV}$).

\section{Observations \label{sec:obs}}
We have determined the light curves of impacts from images obtained by the Atmospheric Imaging Assembly \citep[AIA;][]{Lemen12} onboard SDO in the 1700, 1600, 131, 171, 193, 211~\AA\ channels and thus sample the atmosphere's response over a range of temperatures and hence atmospheric heights.  The  EUV channels were taken with a cadence of 12~s and the UV channels alternated between 1600 and 1700~\AA\ with a 12~s interval between the two channels so the cadence of each UV channel was 24~s.  At the time of the eruption,  304~\AA\ images were taken with a low, 60~s, cadence and were therefore not used in the analysis. 
The 1700~\AA\ emission was mostly continuum from the chromosphere with a small contribution from line emission from the transition region. Similarly continuum from the chromosphere was the main contributor to the 1600~\AA\ channel but there may have been a significant contribution from the transition region  due to \CIV\ emission. The 131~\AA\ emission from the impacts was usually weak and mostly due to \FeVIII.  The 171~\AA\ emission, predominantly from \FeIX, came from slightly cooler plasma (0.63~MK) than the \FeXII\ emission (1.4~MK) seen in the 193~\AA\ channel and the \FeXIV\ (1.9~MK) revealed by the 211~\AA\ channel. Impacts in the chromosphere  probably had additional contributions to the EUV channels from  the transition region \citep{Brosius12} as the plasma was rapidly ionising so temperatures derived from the emission ratios of these bright impacts are likely to be unreliable; however the I$_{171}$/I$_{193}$ or the I$_{171}$/I$_{211}$ should give a good indication of the  temperature of plasma around coronal brightenings caused by impacts.
 
 In all EUV images the cold filament  plasma is seen as dark structures against the background corona due to absorption of H and He photoionising radiation. They are best seen in the well-exposed 171 and 193~\AA\  images. The background is generally smoother in the 193~\AA\ images since it was from the hotter corona, and we used this channel for determining column densities of the impacting fragments. Column densities were derived with the simple one-wavelength method suggested by \citet{Williams13} that uses the relative darkening of the corona due to the filament plasma.  
 
Images from the Extreme Ultraviolet Imager (EUVI) \citep{Wuelser04} at 195~\AA\  on the STEREO-A spacecraft were used in combination with images from  the AIA 193~\AA\ channel to obtain height-time profiles for some of the ejecta. The relatively low cadence of STEREO-A (2.5~min) and the wide separation of the spacecraft (94$^\circ$) made it difficult to unambiguously pick  out the same part of the same ejecta in the AIA/EUVI-A image pairs.
Since the eruption was close to the east limb in the EUVI images, many of the early ejecta falling on the east of the eruption were hidden in the STEREO images due to intervening filament ejecta. The ones best seen from both spacecraft were those from the northern edge and the later, well-separated  ejecta falling from greater heights.

The magnetic fields at the impact sites were obtained from  HMI \citep{Scherrer12} line-of-sight magnetograms. These were  generally weak (<50~G).

\section{Filament eruption} 
\begin{figure}
	\centering
	\includegraphics[width = 0.9\columnwidth]{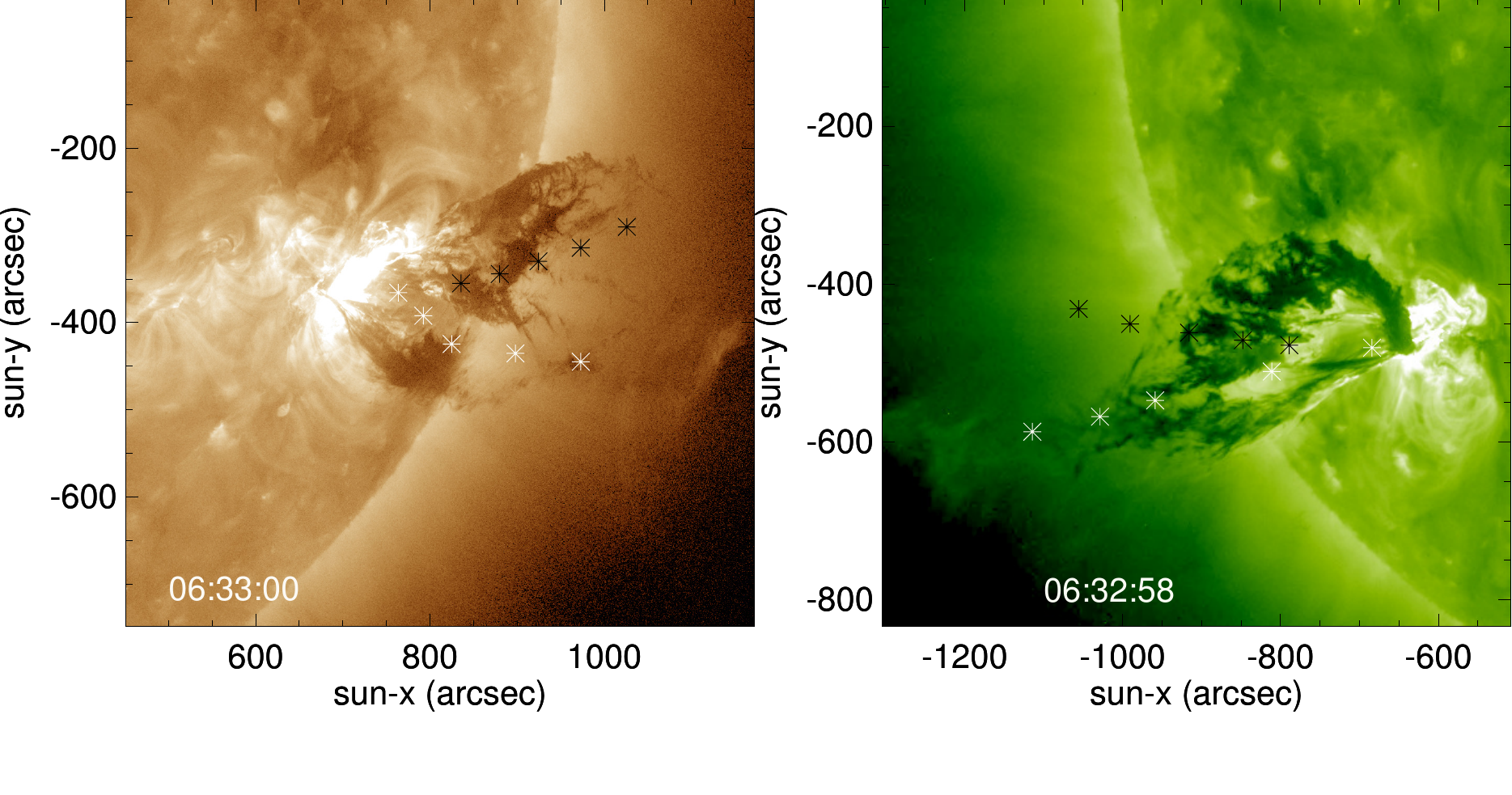}
	\caption{AIA 193~\AA\ and EUVI-A195~\AA\ images of the rising filament plasma. The asterisks mark the paths followed by different parts of the filament. The northern (black) part has a velocity of about 350~\kms, and the southern (white) part about 500~\kms.}
	\label{filament_rise}
\end{figure}

The filament rise started at 06:28~UT. Material was rapidly ejected into the corona with a distribution of velocities so that the central part rose faster and further than the filament sides. Using AIA and EUVI-A images, the paths  taken by different parts of the eruption could be tracked. For example, in Fig.~\ref{filament_rise},
the ejecta along the northern path (black) had an average speed between 320 and 370~\kms, and the speed along the southern path (white)  was between 450 and 600~\kms. The uncertainty is because the ejecta change shape which made it difficult to find exactly the same plasma in consecutive EUVI images, taken 2.5~min apart.

\begin{figure}
	\centering
	\includegraphics[width = 0.9 \columnwidth]{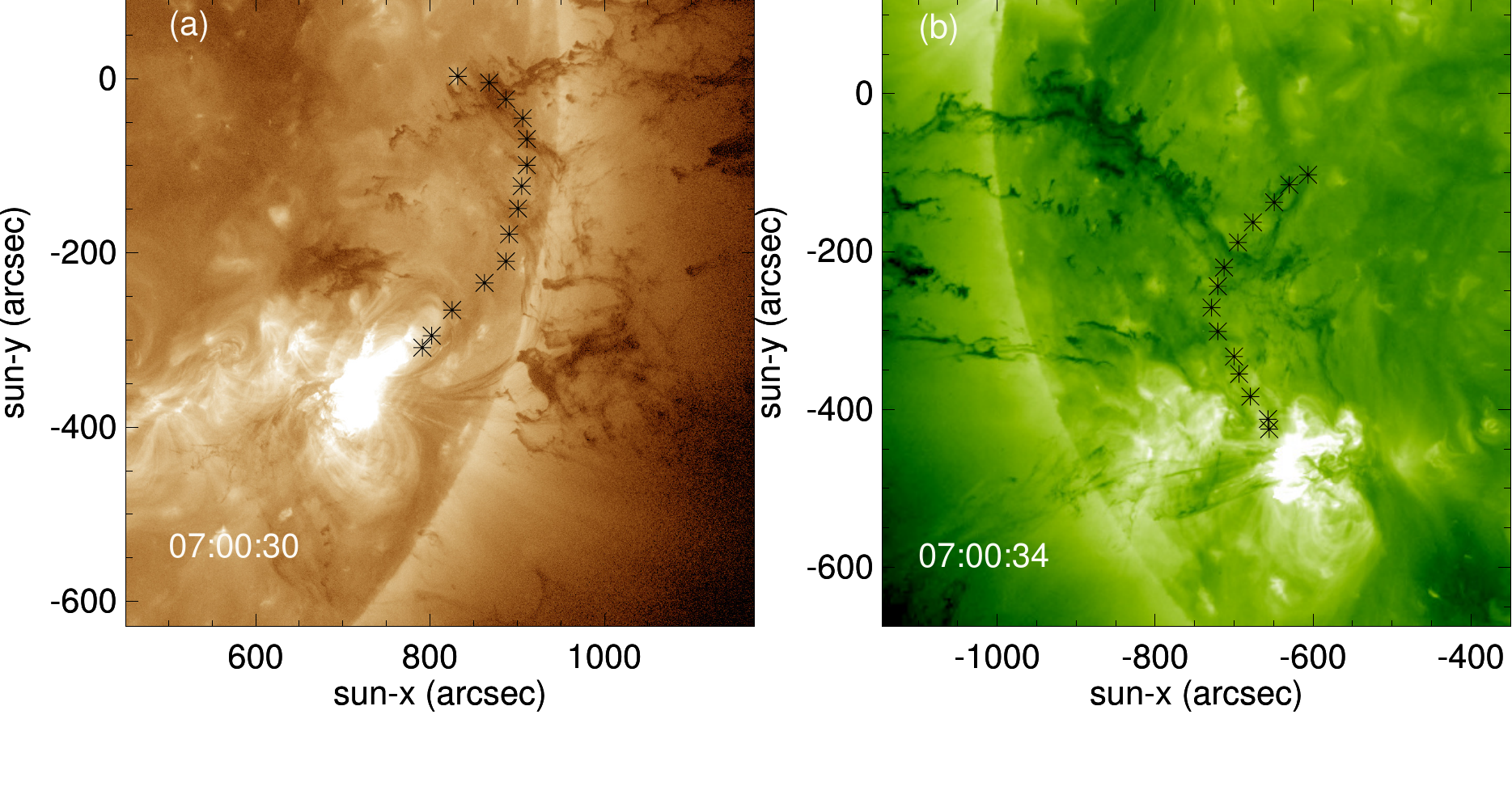}
	\caption{AIA 193~\AA\ and EUVI-A 195~\AA\ images of the descending filament plasma. The asterisks mark the path followed by an early impacting fragment on the northern side of the filament.}
	\label{filament_images}
\end{figure}

\begin{figure}
	\centering
	\includegraphics[width = 0.6 \columnwidth]{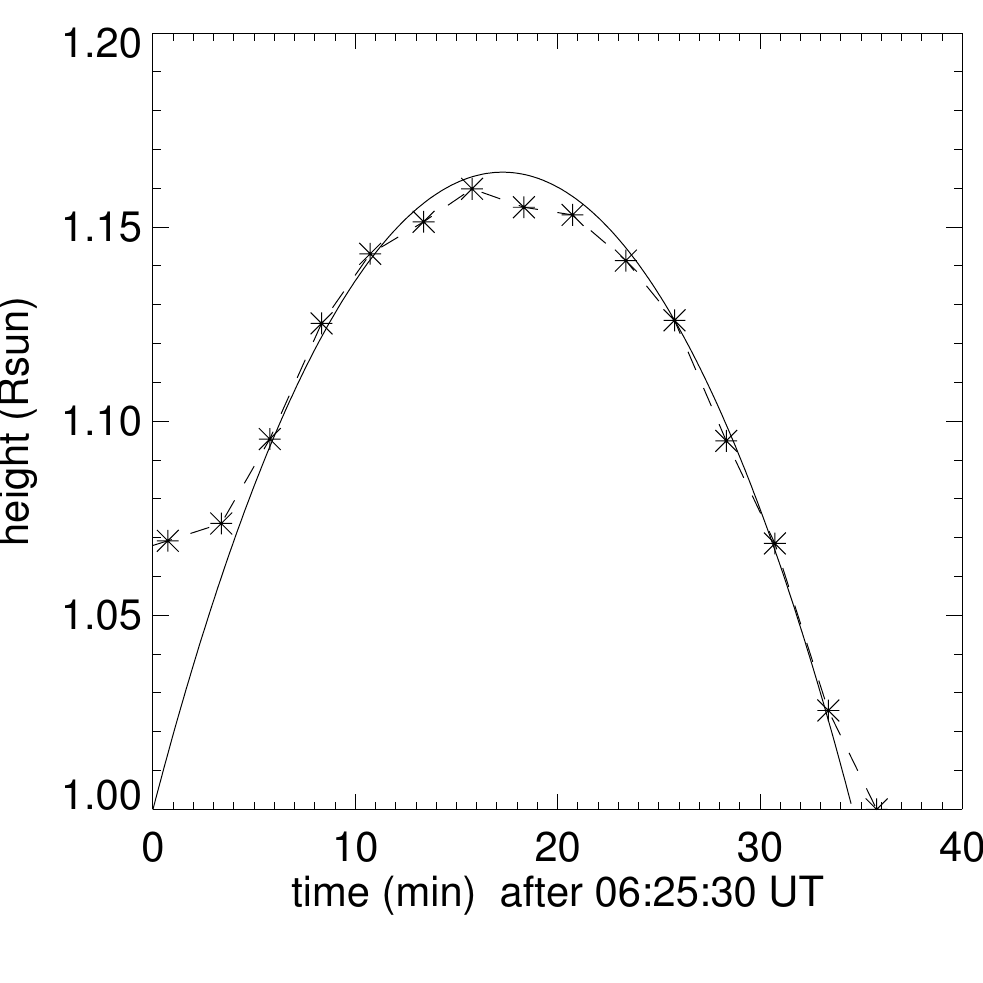}
	\caption{The height-time profile of the early fragment shown in Fig.~\ref{filament_images} (dashed curve) fitted with a parabola representing ejecta from the solar surface with a velocity of 232~\kms, solar  gravitational acceleration of $274~\metss$, and zero drag.}
	\label{filament_path}
\end{figure}

Material from the northern edge is visible throughout its ascent and descent in both AIA and EUVI-A images (Fig.~\ref{filament_images}).  This allowed us to track the height-time profile of an early impact from the northern edge. The height-time path, shown in Fig.~\ref{filament_path},  is well fit by a parabola representing ejecta  leaving the solar surface at 06:25:30~UT with a velocity of 232~\kms. Note that in this fit the ejection from the surface is 2.5~min earlier than the start time of the filament eruption from 1.07~R$_\odot$.
This is the ejecta responsible for `region 3' studied by \citet{Gilbert13}. They used the last two points about 5 min before impact, at 1.03 and 1.07~R$_\odot$, to deduce a velocity of 146~\kms, which they took as the impact velocity. This corresponds with our fit;  however the true impact velocity, obtained by extrapolating to the surface,   is about 80~\kms\ higher. 
In this case, it looks as though  the ejecta did follow a ballistic path but it is the only impact  where we were able to track almost the entire path of the ejecta so we cannot say if other ejecta were also ballistic. Some of the higher ejecta appear to have been slowed  by drag forces \citep{Dolei14}. For all impacts we computed the ballistic impact velocity from the flight time, and for those impacts that could be tracked in both AIA and EUVI-A images, the triangulation velocity about 5 min before impact. 
As discussed by \citet{Reale13} the velocity may be a critical parameter for understanding the impact emission.

\section{Impacts\label{sec:impacts}}

We  selected 16 impacts. Table~\ref{table1} lists the best estimates for the impact landing time, their position, their maximum intensity in the 171~\AA\ channel integrated over a box of roughly 10" x 10", their column density $2-4$~min before impact from the darkening in the 193~\AA\ channel, the mass of the impacting material, the impacts' velocity assuming ballistic motion and an ejection time from the surface at 06:25:30~UT, their velocity measured from triangulation, and the average magnetic field in an area 10\arcsec\ x 10\arcsec\ around the impact.  Triangulation results in a velocity about $50-100$~\kms\ less than ballistic. As mentioned already, the difference can be attributed to the fact that the last triangulation position was measured $2-5$~min before impact and the velocity was computed from the last two points in the trajectory.  The magnetic fields were typically quiet-Sun values except for impact 9 which landed in an active region.  Impact 15 landed in a predominantly negative-polarity coronal hole region.

\begin{table*}
\begin{center}	
	\begin{tabular}{l c c c c c c c c c}	
	\hline
	Impact 	& Time 	&Sun x, y	&I$_{171}$	&Column	 			&Mass		&V (ballist.) 	& V (triang.)	& B+	 	& B- 		\\
			&(UT)	&(arcsec)	&(10$^4$~DN~s$^{-1}$)	& ($10^{18}$cm$^{-2}$)& ($10^{12}$~g)&(\kms)&(\kms)		& (G)	& (G)	\\
	\hline
	1$^1$	& 07:03:05 &820, 30	&77			& 3.6$\pm0.2$		&				&232			& 145		&6 		&5 	  	\\
	 2	& 07:03:53&630, -195&16			& 5.2$\pm0.5$		&				&235			 & 140-180	&5		&6		\\
	3	&07:03:53	&720, -100&13			& 7.7$\pm$0.6		&3-4				&235			 & 165		&3		&3 		\\
	4	&07:10:41&650, 30&6.8			& 5.0$\pm$0.5		&1-2				&265			&150-200		&5		&4		\\
	5	&07:12:17&600, -20	&11			& 3.0$\pm$0.5		&				&275			&			&5		&4		\\
	6	&07:16:17	&735, 175	&4.8			&3.2$\pm$0.1		&				&290			&			&6		&6		\\
	7	&07:17:29	&735, 145	&2.8			&3.0$\pm$0.1		&				&290			&			&4		&6		\\
	8	&07:21:05&485, -100&5.9			&3.6$\pm$0.3		&				&305			&			&5		&7		\\
	9	&07:23:29	&620, 245&89			&6.6$\pm$0.4		&2-3				&310			&250			&6		&30		\\
	10	&07:26:17&415, -75	&5.0			&3.0$\pm$0.4		&				&315			&			&5		&4		\\
	11$^2$	&07:28:17	&520, -80	&21			&2.6$\pm$0.3		&				&320			&			&6		&6		\\
	12		&08:07:53	&405, -75	&66			&3.7$\pm$0.1		&				&400			&300			&6		&5		\\
	13$^1$	&08:08:41	&390, -65	&220			&3.7$\pm$0.1		&2-3				&405			&300			&6		&5		\\
	14	&08:50:48	&300, 40	&31			&3.9$\pm$0.2		&2-3				&445			&380			&4		&4		\\
	15	&08:51:31	&315, 120	&49			&<1.5			&				&445			&380			&3		&7		\\
	16	&08:51:55	&415, 65	&10			&2.2$\pm$0.2		&				&450			&380			&3		&3		\\
	\hline
	\end{tabular}
	\caption{Parameters of the selected impacts. I$_{171}$  are the peak impact 171~\AA\ channel intensities. The column densities are an estimate of the neutral hydrogen component 2-4~min  before impact. The mass was computed by integrating the column densities in the falling plasma. The ballistic velocity assumes an ejection time from the surface 06:25:30~UT. The triangulation velocity is the velocity about 5~min before impact. The magnetic field strengths, B+ and B-, are the average positive and negative field  in the region around the impact.
Marked impacts have been analysed by $^1$\citet{Gilbert13}, $^2$\citet{Reale14}.
\label{table1}}
\end{center}	
\end{table*}

\subsection{UV and EUV intensities}
 \citet{Gilbert13} have shown that on impact there was generally a rapid increase to peak intensity in all EUV channels simultaneously, followed by a slower intensity decrease. The sharp rise was caused by the initial impact.  The increase in pressure due to the impact subsequently changed the environment into which the trailing fragment plasma was falling. In several events bright EUV emission, looking like a splash, was seen propagating back from the impact site. The characteristics of one of the largest splashes is described later.  In this section we concentrate on emission from the bright impacts, not their surroundings. 

To avoid emission from the surroundings, we have chosen compact, short-lived impacts, and obtained light curves in the close vicinity of the impacts.  We typically used boxes of size 10" x 10" but they varied according to the size of the bright emission kernel. When the impact was part of a larger group, the  intensities were taken from a small region centred on the brightest 1600~\AA\ site. If there was no noticeable 1600~\AA\ brightening then the region is centred on the brightest 171~\AA\ patch.  
Since the impacts were much brighter than the surroundings and were computed after background subtraction, results are not sensitive to the box size. 
   
\begin{figure*}
	\centering
	\includegraphics[width = 0.9\linewidth]{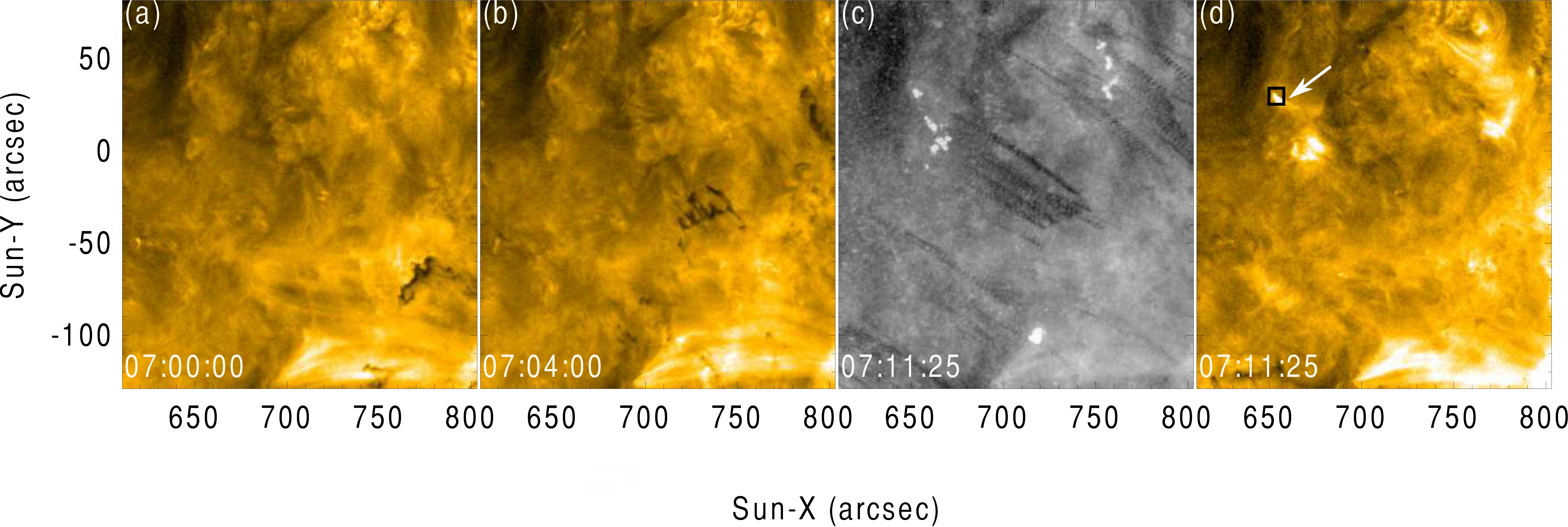}
	\caption{The relationship between the downflowing plasma and the impacts at 1600~\AA: (a, b) AIA 171~\AA\ images showing the finger and arc structure of the downflowing material; (c)  composite of dark tracks of 193~\AA\ absorption in the fragment tips and the bright 1600~\AA\ impact emission; (d) the impact in 171~\AA. Dense structures in the downflowing plasma impact the chromosphere resulting in circular brightenings in 1600~\AA\ (c). In (d) the arrow points to impact 4 and the black box shows the region used for integrating the fluxes. Panel (c) is the central impact in the movie composite\_1600. The region is indicated by an arrow at the end of the movie.}
	\label{composite}
\end{figure*}

Fig.~\ref{composite} shows an example of a fragment group. The group started in the corona as a single finger, the front then broke up into at least 6 smaller fingers. The physics of the break up is still not well understood. It seems to have been influenced by the magnetic field and the interaction of the wind with the partially ionised falling plasma \citep{Innes12}. Each high density finger tip led to a bright circular impact at 1600~\AA.  This is illustrated in Fig.~\ref{composite}(c) where we show a composite of the brightest 1600~\AA\ and the dark,193~\AA\ downflow tracks. Downflow tracks were used by \citet{Reale13} to show the motion of the falling fragments. These tracks are given by the minimum intensity value of each pixel during the downflow. 
The 171~\AA\ emission from the impact was more diffuse, covered a larger area and lasted longer than the 1600~\AA. It came both from the corona and the impact site. 
The impact, indicated by an arrow and surrounded by a box in (d), is impact 4. The box indicates the region over which the intensities were integrated.

As seen in Table~1, the impacts' 171~\AA\ intensities covered a range  of almost two orders of magnitude and show no simple relation  to either the mass or velocity in the falling plasma.  
The intensities are integrated over the box surrounding the impact site and their intensity depends more on the size of the impact site, than on the maximum intensity over the impact site. 
For example impact 13 was seven times brighter than  14 although it had the same mass and was slower.  As discussed below, the enhanced EUV intensities were probably due to dissipation of kinetic energy in the transition region or corona  where trapping along low-lying coronal field prevented the fragment from falling as deep in the chromosphere as the HD models of  \citet{Reale13} predict.

\begin{figure*}
	\centering
	\includegraphics[width = 1 \linewidth]{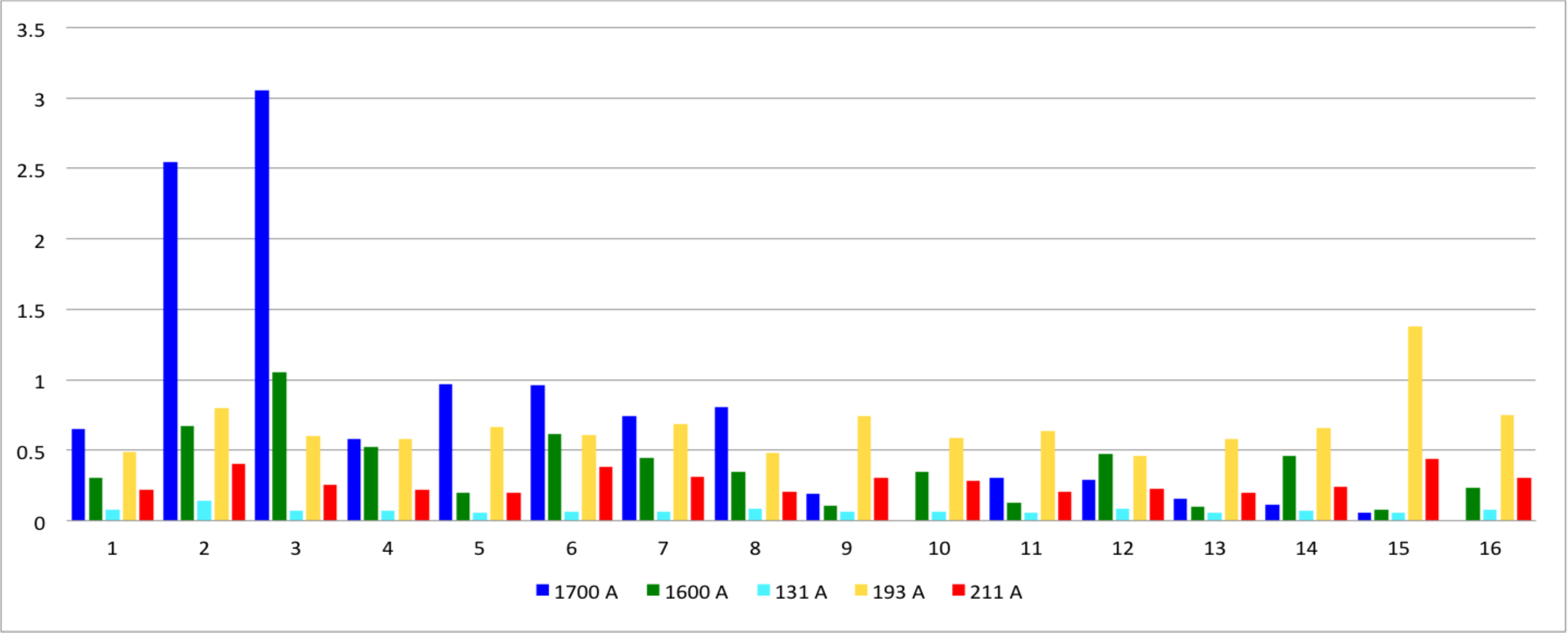}
	\caption{Impact intensity in different UV and EUV channels relative to the 171~\AA\ intensity. Note the decrease in I$_{UV}$/I$_{EUV}$  with time and the increase in I$_{193}$/I$_{171}$  for later impacts.  }
	\label{intensities_fig}
\end{figure*}

\begin{figure}
	\centering
	\includegraphics[width = 1.0\columnwidth]{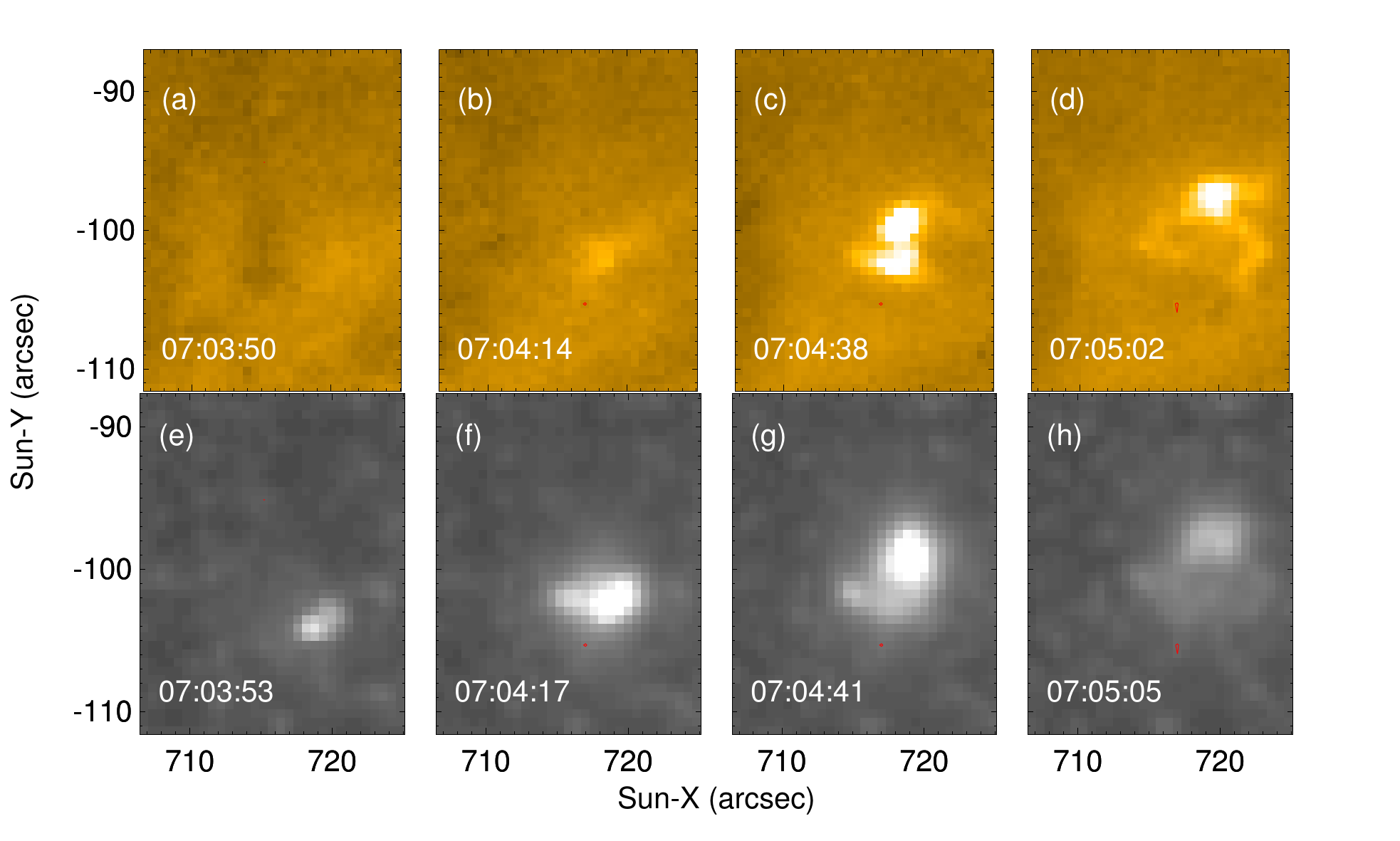}
	\caption{The early low-velocity impact 3 with strong UV to EUV: top row 171~\AA: bottom row 1600~\AA.}
	\label{dyrus_fig}
\end{figure}

The maximum of the impact's intensity in several channels relative to the maximum  171~\AA\ intensities are represented as a column chart in Fig.~\ref{intensities_fig}.
Two early impacts, 2 and 3, stand out because of their much higher-than-average I$_{UV}$/I$_{EUV}$. Both these impacts had an impact velocity between 230 and 250~\kms\ and landed in quiet-Sun. 
A close up of impact 3's 1600 and 171~\AA\ emissions is shown in Fig.~\ref{dyrus_fig}. The 171~\AA\ brightening occurred about 30~s after the 1600~\AA\ brightening. 
According to the HD simulations of \citet{Reale13}, impacts with velocity about 250~\kms\ should not produce significant EUV emission so there must be additional dissipation effects not accounted for in their simulations.  

\begin{figure}
	\centering
	\includegraphics[width = 0.5\columnwidth]{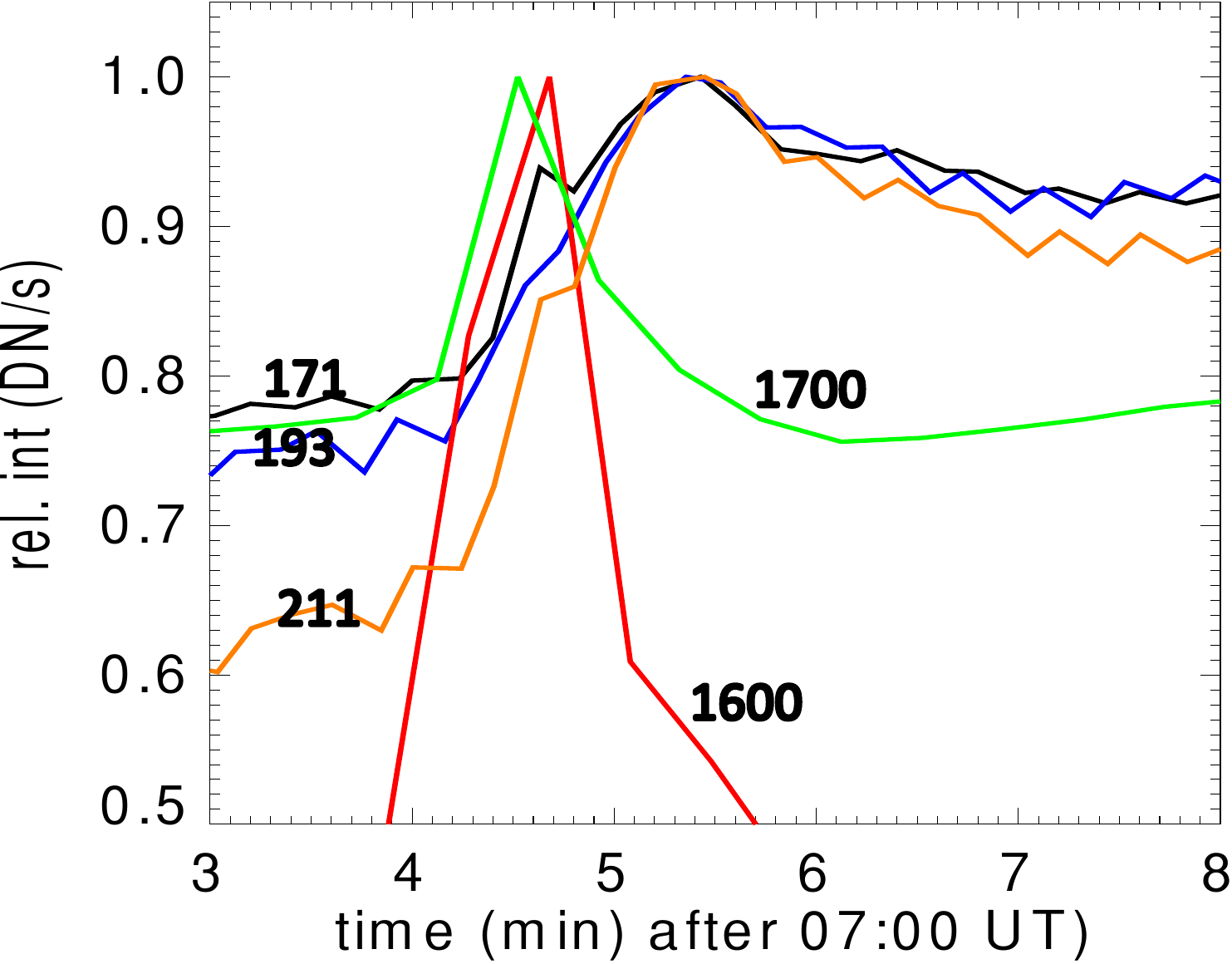}
	\caption{UV and EUV light curves of the early low-velocity impact, impact 3, shown in Fig.~\ref{dyrus_fig}. The curves are labelled with their channel wavelength in \AA. }
	\label{lc_dyrus_fig}
\end{figure}

The time delay was typical  for the EUV and UV emissions, as noted by \citet{Reale14}. It was seen in all impacts with roughly circular UV emission kernels. An explanation, given in \citet{Reale14}, is that the 1600~\AA\ is from a second stream of falling fragments that is shock-heated to transition region temperatures by the backflow from the initial impact; however in most of the impacts studied here (e.g. Fig.~\ref{composite}) the plasma accumulates at the head of the fragment and the 1600~\AA\ appeared in front of the downflow tracks so this explanation is unlikely. Quantitative computations showing the piling up of plasma at the head of fragments have been reported by \citet{Carlyle14}. Another problem with the \citet{Reale14} scenario is  that the main peaks in the 1600~\AA\ are almost 100~s after the initial impact (Fig. 1c in \citet{Reale14}) but the observations show only a single peak  from the impact site. 
 Also the brightening was seen from the same sites and with similar light curves in both the 1600 and 1700~\AA\ channels (Fig.~\ref{lc_dyrus_fig}), implying that it is chromospheric not transition region emission. 
To understand the delay between the EUV and UV requires more detailed models of the dynamics and ionization of the chromosphere's reaction to the impacts, including magnetic effects.

The circular shape of the 1600~\AA\ suggests that the emission was from the essentially uniform chromosphere rather than the more structured corona. Many impacts produced thread-like structures at 171~\AA\ surrounding or connecting the main impact site suggesting that although quiet Sun, the corona may still be  structured by the magnetic field. 

There were two impacts with abnormally low I$_{UV}$/I$_{EUV}$; the one  in the active region (9), and one low column-density impact (15). The low column-density impact  (15) was the only one with  a I$_{193}$/I$_{171}$ greater than one. The ratio was around 0.5 for most other impacts.
 
\subsection{Column densities}
Since the kinetic energy of the impact is critical to understand the emission, the impacting mass is an important parameter. We have tested various methods to calculate the mass but are not satisfied with any. The computation of mass is subjective because as the  plasma fell it broke up creating a broad front of fragments (e.g. Fig.~\ref{composite}) so that one needs to decide when to measure the mass. Also if the falling plasma was in the form of long fingers, it was not obvious which part to take as  the impact mass. In several cases, however, it was possible to identify the impacting mass. The estimates given in Table~1 were obtained in these cases by integrating the hydrogen column density of impacting plasma over the fragment area, dividing by the hydrogen abundance, 0.92, and multiplying by the pixel size, 0.19 Mm$^2$, and the mean atomic mass, $2\times10^{-24}$~g.

\begin{figure}
	\centering
	\includegraphics[width = 0.7\linewidth]{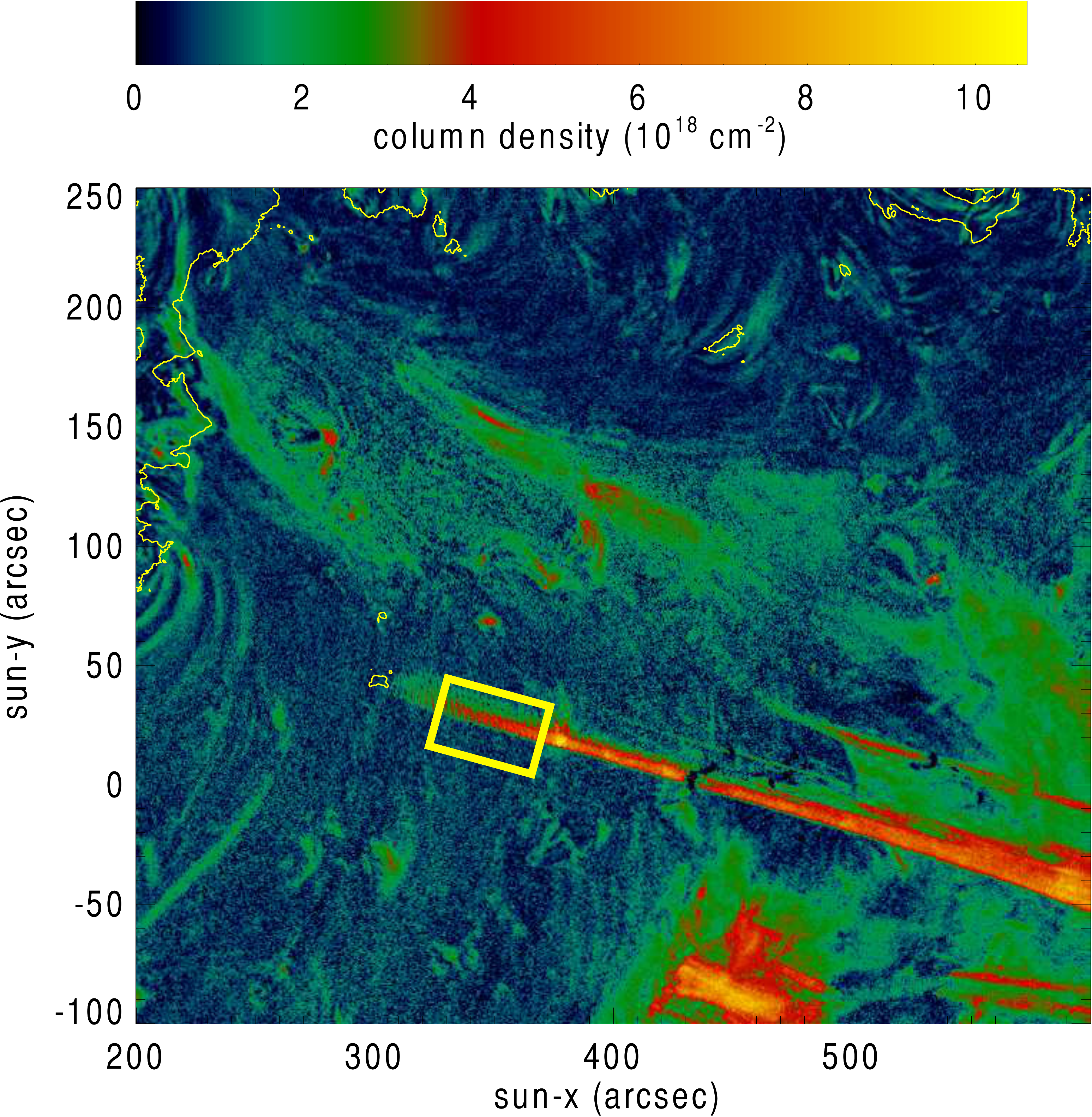}
	\caption{Column density during the downflow of impact 14. The yellow box outlines the region used to compute the column density in Table 1.}
	\label{columndensity}
\end{figure}

For all impacts we were able to estimate the neutral hydrogen column density in individual fragments.
It was not possible to determine the line-of-sight width of the fragments so densities could not be calculated,  nevertheless column density may give a rough idea of the relative densities of the fragments.  The column densities were obtained  by taking the ratio of the track image to a background image in the 193~\AA\  channel which gives the evolution of the maximum optical depth at 193~\AA\ in the falling fragment. Then using Equation 4 and Figure 2(a) of \citet{Williams13}, the evolution of the neutral hydrogen column density was calculated. The column density, given in Table~1, is the average hydrogen column density of the highest 150 pixels  2-4 min before impact. The uncertainty  is the corresponding variance. In Fig.~\ref{columndensity}, we show the column-density track of impact 14. A yellow box shows the region from which the highest 150 column densities were selected to compute the column density for the fragment. The actual number of pixels taken was not critical, and we found similar column densities for 30\% more or less pixels.  We note that close to the solar surface the neutral hydrogen column density decreases by at least a factor two, implying ionisation before impact. The ions will be trapped by the coronal magnetic field causing bright EUV emission outside the main impact site (see movie\_impact15).

The range of column densities, a factor four, is much smaller than the range of 171~\AA\ intensities. It is probably not the main parameter in determining the 171~\AA\ intensity but as discussed below it may play a factor in determining I$_{UV}$/I$_{EUV}$.

\subsection{Magnetic field}
We have looked at the relationship of the impacts' intensities to the local field at the impact site. The majority of impacts occurred in quiet Sun and we include one impact on the edge of an active region  (9) and one low column-density impact  in a coronal hole region (15). 

\begin{figure}
	\centering
	\includegraphics[width = 0.9\linewidth]{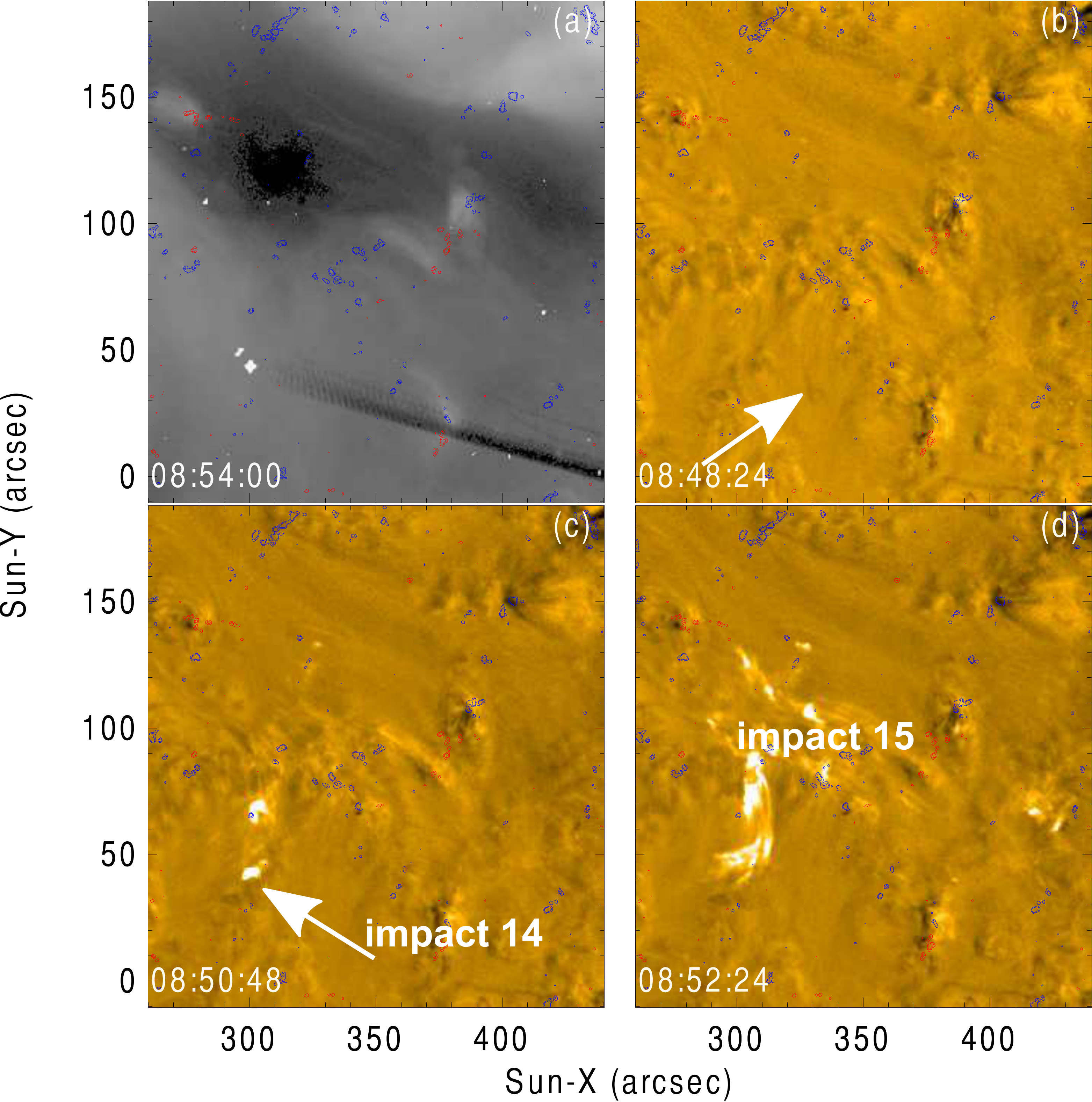}
	\caption{Overview of impacts 14 and 15, as seen in UV  (1600~\AA) and EUV  (171~\AA) channels. The red/blue contours outline magnetic fields of +/-50~G. (a) composite of 1600~\AA\ and downflow tracks. (b) 171~\AA\ base-difference image showing a faint downflow. (c - d)  171~\AA\ base-difference images of the impact, scaled between $\pm300$~\DNS.  The white arrow points to the position of a very faint falling fragment. It is visible in the movie, impact15\_movie. Impact 14 is the compact site indicated by an arrow in (c). Impact 15 is the large region of diffuse 171 in (d). 
}	 
	\label{gbm}
\end{figure}

The images of impacts towards the coronal hole show bright EUV threads that were probably illuminating the structure of the magnetic field on the edge of the coronal hole.  They are best seen in the movie, impact15\_movie. Here one sees one dark fragment to the south and a broad, very faint fragment heading towards the coronal hole. Fig.~\ref{gbm} gives an overview of the downflows and impacts. The downflow track and 1600~\AA\ emission in Fig.~\ref{gbm}(a) and the southern 171~\AA\ emission in (c) is impact 14.  The display of brightenings on the edge of the coronal hole seen in (d) are grouped together as impact 15. They were produced by a faint sheet-like fragment that is not visible in individual images but can be seen in impact15\_movie.
Although there were no high column-density fragments associated with the EUV threads, they were   as bright in the  EUV  as the neighbouring  impact caused by a high column-density impact in the chromosphere.

\begin{figure}
	\centering 
	\includegraphics[width = 0.9\columnwidth]{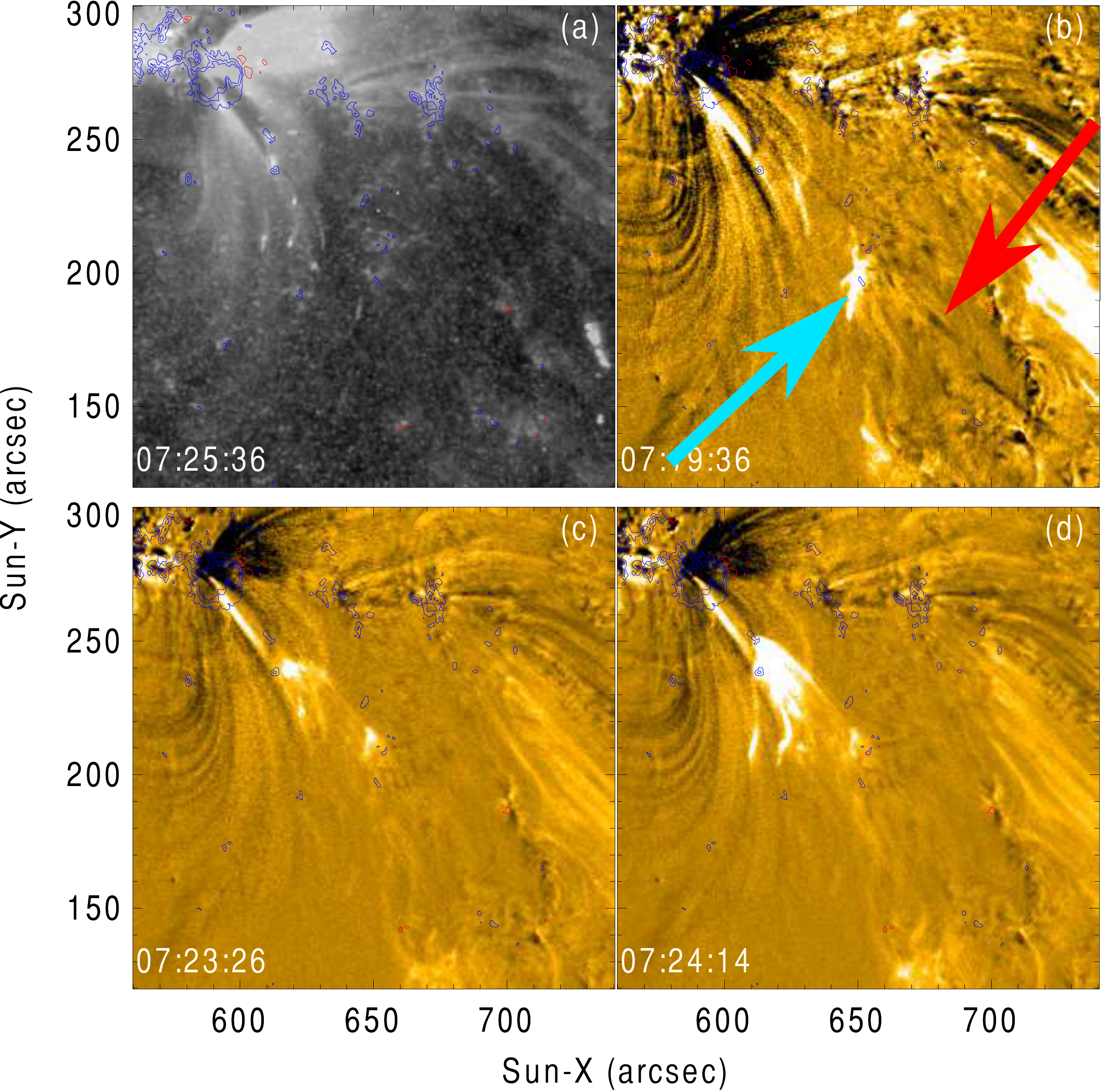}
	\caption{Overview of an impact, impact 9, in an active region as seen in UV (1600~\AA) and EUV (171~\AA) channels. The red/blue contours outline magnetic fields of +/-100, 300, 500~G. (a) composite of 1600~\AA\ and downflow tracks. (b) 171~\AA\ base difference image showing the downflow filaments. (c - d)  171~\AA\ base-difference images of the impact, scaled between $\pm300$~\DNS. The base image was at 07:10:00~UT. The red arrow points to the impacting filament and the white arrow points to an earlier impact that seems to have affected the structure of this impact.
	The evolution of the impact is shown in impact9\_movie.
	}
	\label{piglet}
\end{figure}

The impact in the active region, impact 9, was similar to the low column-density impact in that it produced lower I$_{UV}$/I$_{EUV}$ for its velocity than impacts in the quiet-Sun chromosphere. The accompanying movie of the impact, impact9\_movie, shows that it started as a high column-density fragment that, about 3~min before impact, dispersed in the corona as it crossed a region affected by an earlier small  splash. An overview of the impact is provided by  Fig.~\ref{piglet}. 
The images show a bright fan-shaped region of EUV emission at the impact site and very little UV, implying that the downfalling fragment dissipated its kinetic energy in the corona  and did not  reach the chromosphere.


\subsection{Splashes}
\begin{figure*}
	\centering
	\includegraphics[width = 1.0\linewidth]{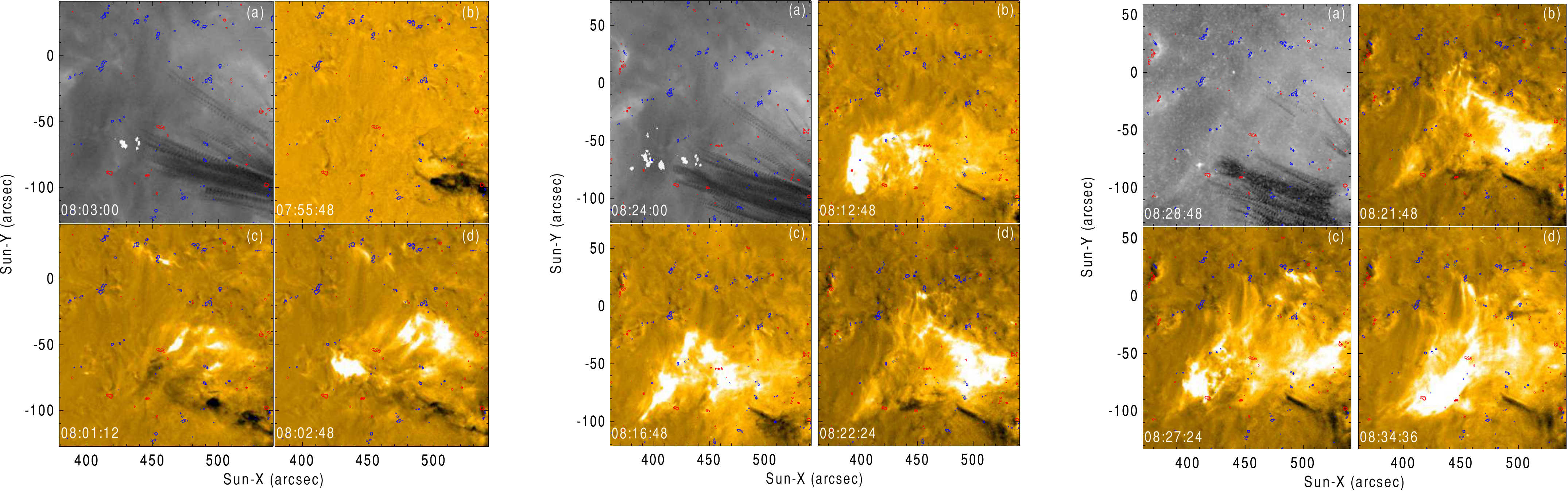}
	\caption{Overview of the two consecutive streams in the large impact, impact 12, as seen in UV (1600~\AA) and EUV  (171~\AA) channels.
	Each panel of four shows in (a) a composite of 1600~\AA\ emission and tracks of downflows from 193~\AA\ images and (b-d) base-difference images chosen to show the movement of the bright splash emission. The red/blue contours outline magnetic fields of +/-50~G. The base image was taken at 7:50:00~UT. The 171~\AA\ images are scaled between $\pm300$~\DNS. Left: The first EUV brightenings were from low density impacts that dissipated in the corona.  Middle: A strong splash was created by the first series of  chromospheric  impacts. Right: The splash from the first series prevented the second series of fragments from reaching the chromosphere so there was very little 1600~\AA\ emission. Movies impact12\_171movie and impact12\_193movie, available online, show the full sequence of base-difference images at 171 and 193~\AA.}
	\label{imp_3}
\end{figure*}

EUV splashes were seen after several of the impacts in the  quiet Sun. They mostly appeared where the impacting fragment was unusually long or wide. Several of the  splashes were as bright as the impact itself in the 171~\AA\ channel. 
The  main features of quiet-Sun splashes are illustrated by the biggest sequence of impacts that started  around 08:00~UT and had an impact speed about 400~\kms. As shown in impact12\_movie, there were two successive streams of impacts following roughly the same path. The first stream, shown in the right and middle panels of Fig.~\ref{imp_3}, produced the `region 1' investigated by \citet{Gilbert13}. Where the impact was in the chromosphere, the landing site brightened in all channels (UV and EUV). The splash, represented in the lower images of the middle panel, moved back towards the inflowing material with a plane-of-sky velocity of about 100~\kms. The effect of the splash on the second stream was to prevent the impacting fragment from reaching the chromosphere. As shown in the top-left image of the the right-hand panel of Fig.~\ref{imp_3},  little UV emission was produced by the second stream. 
 For comparison, the splash structure simulated by \citet{Reale14} had a speed of 200~\kms\ and was more extended in the direction away from the infalling plasma.

\begin{figure*}
	\centering
	\includegraphics[width = 0.7\linewidth]{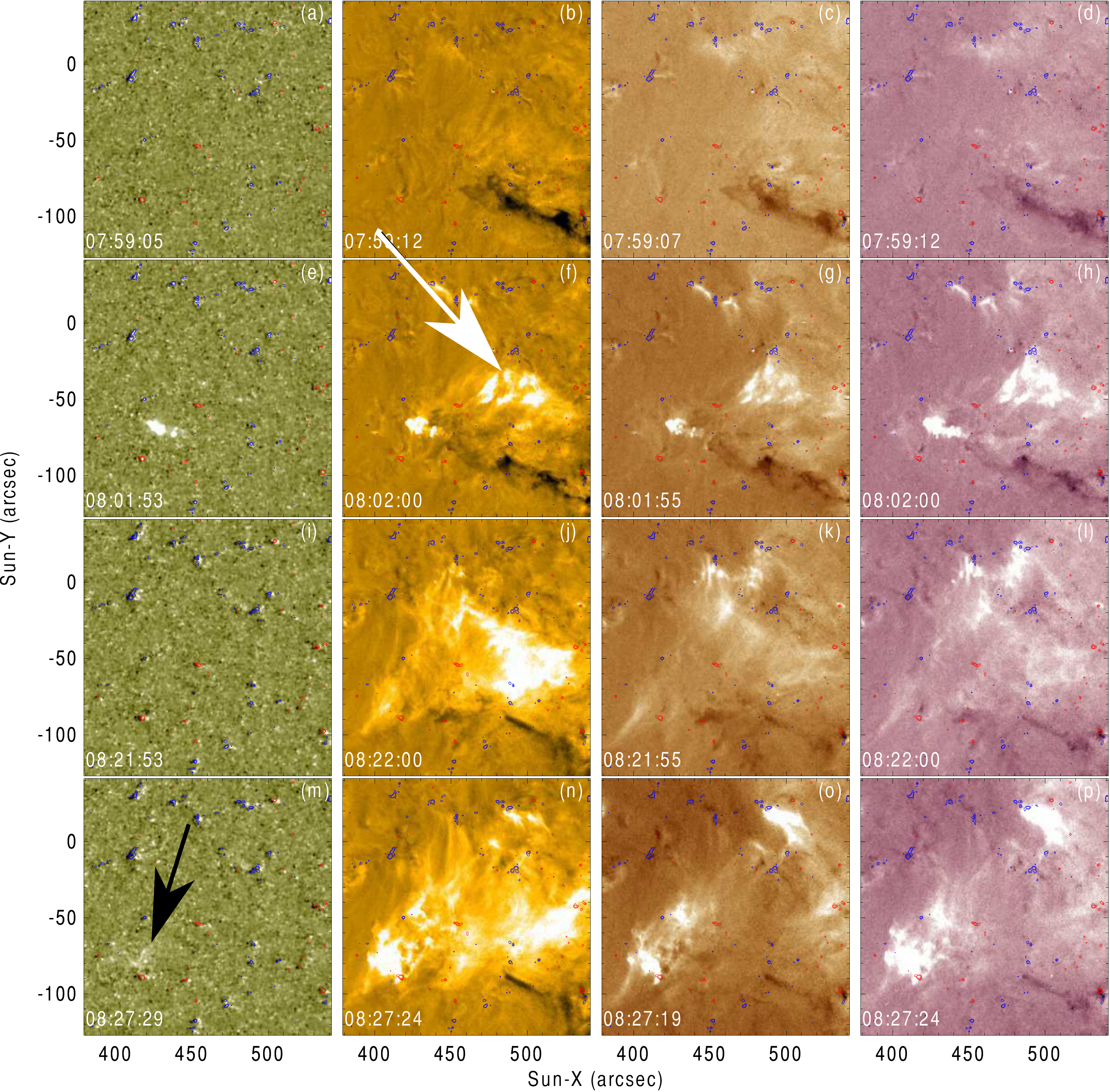}
	\caption{Overview of the large impact, impact 12, as seen in UV (1600~\AA) and EUV  (171,  193 and 211~\AA) channels. The EUV images are base differences with a base image at 07:50:00~UT. The red/blue contours outline magnetic fields of +/-50~G. The white arrow in (f) points to bright EUV emission produced by a low density impact. The black arrow in (m) points to diffuse 1600~\AA\ emission during the second stream of impacts.
	}
	\label{imp_211}
\end{figure*}

Fig.~\ref{imp_211} shows the impact and splash sequence in the 1600, 171, 193 and 211~\AA\ channels.
The lower density, filamentary parts of the downflows on the edge of the main downflow, dissipated in the lower corona producing bright EUV emission (e.g. the region in (f) marked with a white arrow). The splash appeared later and was visible in the EUV channels. Compared with the neighbouring low column-density impacts dissipating in the corona, the splash had a higher I$_{171}$/I$_{193 }$  (or I$_{171}$/I$_{211}$). This implies that the splash was cooler than the impacts dissipating in the corona. 

\begin{figure}
	\centering
	\includegraphics[width = 0.7\linewidth]{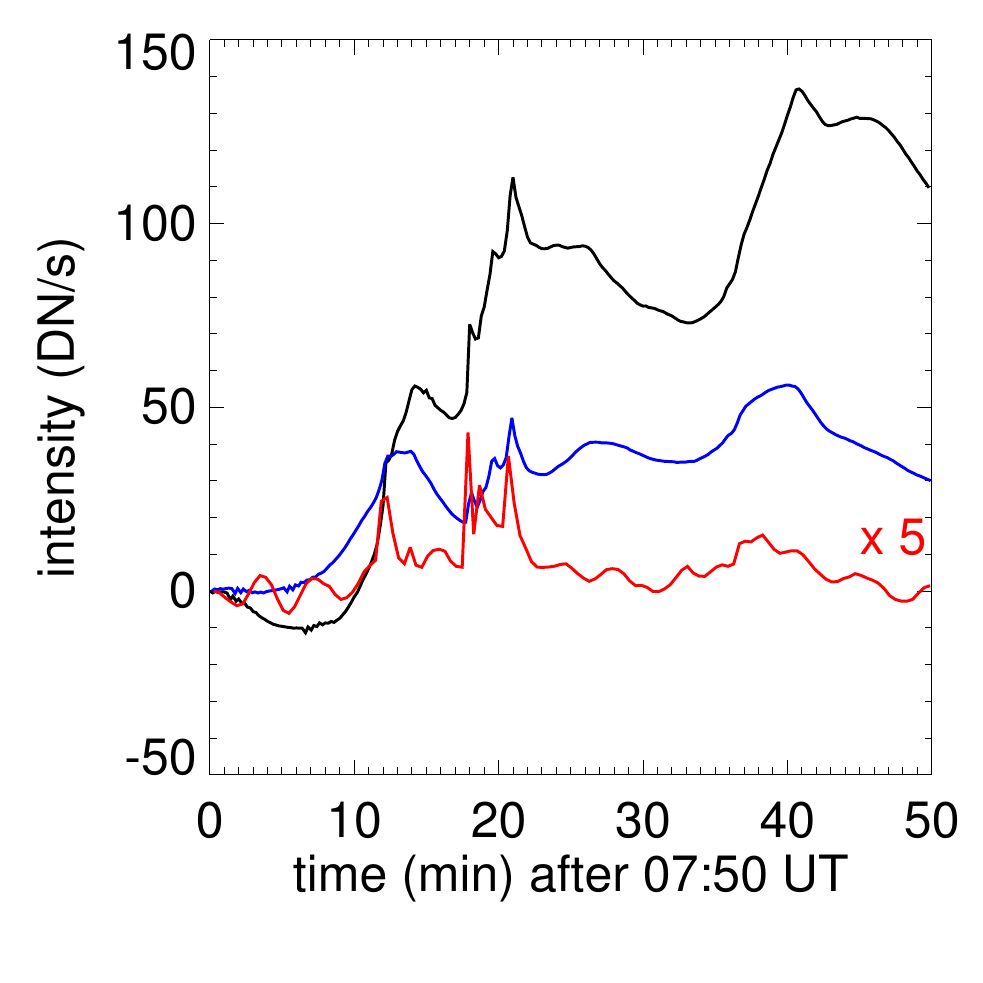}
	\caption{Light curves during the big impact, 12,  taken over the region shown in Fig.~\ref{imp_211}: red -  1600~\AA, black -171~\AA, and blue - 193~\AA. }
	\label{lc_imp}
\end{figure}

The light curves of the 1600,  and 193~\AA\ emission, obtained by integrating  base-difference intensities over the whole region, are  shown in Fig.~\ref{lc_imp}. 
The head of the fragment produced the brightening around 12~min seen in the 1600~\AA\ image (Fig.~\ref{imp_211}(e)). The impacts 12 and 13 caused the series of spiky brightenings between 17 and 22~min. A broader, more gradual  increase of 1600~\AA\ emission occurred around 37~min. 
The images show that this is because there were many small, barely perceptible  brightenings at 1600~\AA\  (Fig.~\ref{imp_211}(m)) coinciding with the EUV impact emission from the second stream.  
We note that there was no bright 1600~\AA\ emission in the corona contrary to the predictions by \citet{Reale14}. This may be because the density of the observed fragments was  lower than those simulated by \citet{Reale14}. They assumed spherical fragments with density $5\times10^{10}$~\cc\ and radii $1.4-2.6\times 10^8$~cm , whereas the dark fragments in Fig~\ref{imp_211} have widths of typically $8\times10^8$~cm and column densities $3.7\times10^{18}$~\cmsq\ which corresponds to a density of about $5\times10^9$~\cc. The observed fragments  may therefore have had a factor ten or so lower density than the simulated ones, and this could explain why the fragments rapidly ionise in the corona and produce very little 1600~\AA\ emission.

There are subtle differences in the 171 and 193~\AA\ intensities that can be attributed to the effects of the low column-density impacts and the splashes. The low  column-density impacts caused the early increase in the 193~\AA\ intensity and the bump in the 193~\AA\ intensity around 25~min. 
The splash was responsible for the increased 171~\AA\ intensity immediately after the impacts at 20 and 40~min. 
The decrease in 171~\AA\ intensity at the beginning is because the falling fragment obscured bright 171~\AA\ emission during its descent.

\begin{figure*}
	\centering
	\includegraphics[width = 0.8\linewidth]{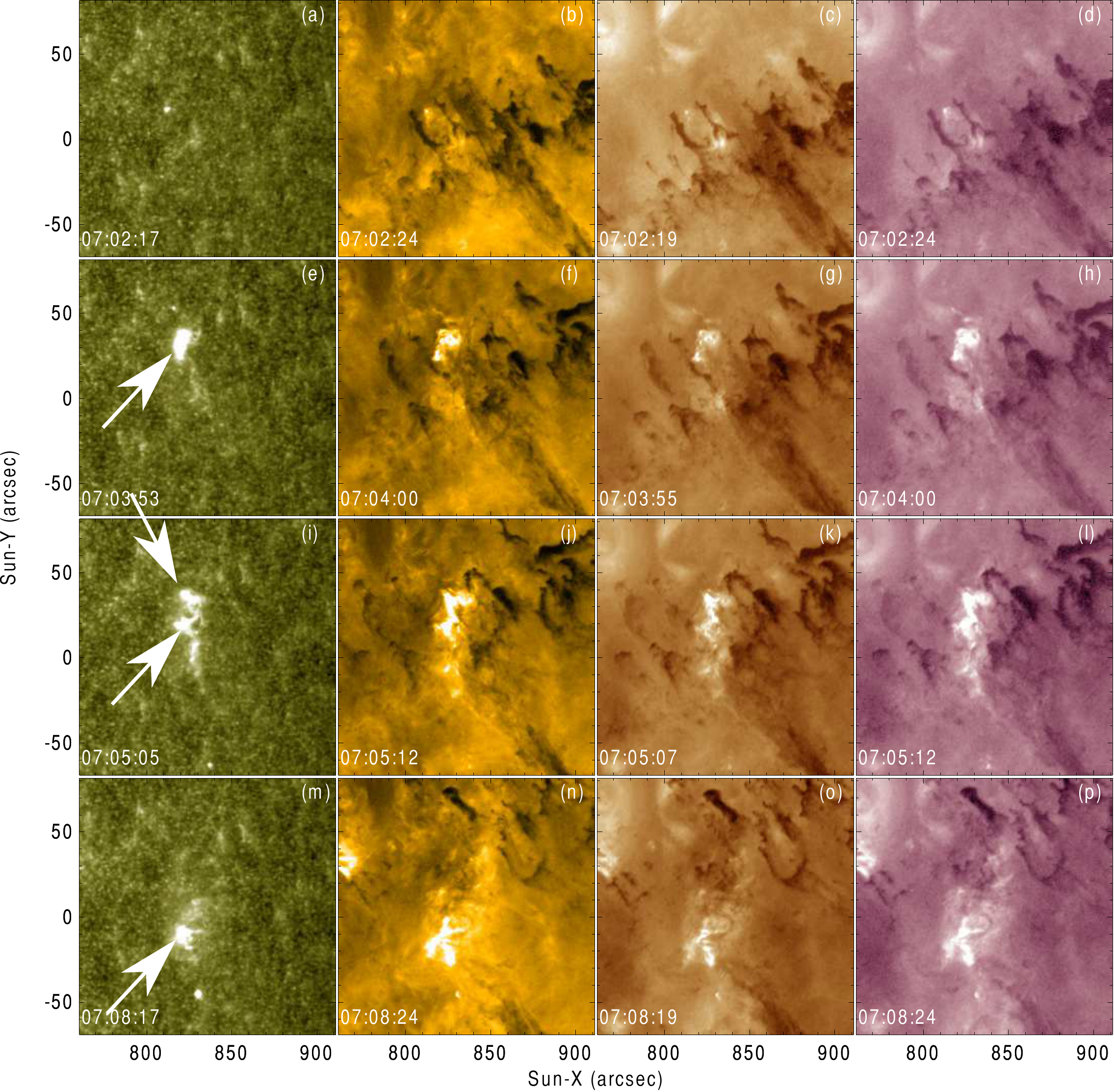}
	\caption{Overview of a large early impact just after '1' as seen in UV (1600~\AA) and EUV  (171,  193 and 211~\AA) channels. The EUV images are base differences with a base image at 07:50:00~UT. The movie impact1\_movie shows the evolution at 171~\AA. The arrows point to 1600~\AA\ sites that were also bright at 1700~\AA.
	}
	\label{aty_211}
\end{figure*}

A second large impact with interesting splash effects in the 1600~\AA\ and various EUV channels is shown in Fig.~\ref{aty_211}. This  was the only impact in our study
with structured 1600~\AA\ suggesting bright \CIV\ emission from falling material above the chromosphere especially as it was seen along
the front of a second stream of infalling plasma   \citep{Reale14}. The brightest sites, indicated with white arrows in Fig.~\ref{aty_211}, were probably chromospheric impacts because they produced circular brightenings at 1700~\AA\ as well. The threads in between have much lower I$_{1700}$/I$_{1600}$, indicating  transition region emission. 
These threads are unlikely to have been caused by scattered impacts in the chromosphere.
It was also the lowest velocity impact. Thus consistent with  high I$_{UV}$/I$_{EUV}$ observed in low-velocity impacts in the quiet chromosphere we see that low-velocity impacts dissipating higher in the atmosphere also have higher I$_{UV}$/I$_{EUV}$. The high I$_{211}$/I$_{171}$ at the same site suggests that the 1600~\AA\ was from the impacting plasma not from the splash.
The splash can be seen in impact1\_movie propagating back from the impact site in the direction of the falling material.

\section{Discussion}
We have studied the UV and EUV emission of 16 filament fragments that fell onto the Sun after the 2011 June 7 eruption with impact velocities ranging from 230 - 450~\kms\ assuming  ballistic propagation: an assumption that has been justified by stereoscopically tracking the path of one fragment in EUVI-A and AIA images. The impacts range  over two orders of magnitude in 171~\AA\ intensity,  and had column densities ranging from less than 1.5 to 7.7$\times10^{18}$~cm$^{-2}$. All impacts produced bright EUV emission and the strength of the I$_{UV}$/I$_{EUV}$ emission depended on impact velocity, column density, magnetic environment, and presence of splashes caused by earlier impacts.

For impacts in the quiet Sun not affected by splashes, the I$_{UV}$/I$_{EUV}$  decreased with velocity. 
This suggests that at low velocities there was not as much impact energy to heat and ionise all plasma at the impact site to coronal temperatures so the I$_{UV}$/I$_{EUV}$ was relatively large \citep{Sacco10}. The lowest column-density impact had the highest I$_{193}$/I$_{171}$ suggesting that low density impacts result in higher temperature emission. Image sequences of the  plasma during the largest quiet-Sun impact (Fig.~\ref{imp_211}) showed the same tendency. We suspect that the reduced I$_{UV}$/I$_{EUV}$ was because low density filaments become essentially fully ionised before reaching the chromosphere. Therefore the impacting plasma was trapped in the corona and did not reach the chromosphere. Since the coronal density was low, the impact rapidly heated the  coronal plasma to temperatures above 2~MK, and conduction transported the heat down to the chromosphere. In contrast, the kinetic energy of  impacts  into the high-density, low-temperature  chromosphere  was concentrated at the impact site creating a large increase in local pressure \citep{Reale13}. The dissipated energy heated the local chromosphere from less than $0.01$ to more than 3~MK, resulting in a broad range of emission temperatures, with a bias towards the cooler coronal values. 

Dissipation in the corona also explains the emission observed from impact 9 in the active region. Compared to impacts with the same velocity in the quiet Sun, the I$_{193}$/I$_{171}$ was higher and the I$_{UV}$/I$_{EUV}$ lower. 
Also one sees in
  impact9\_movie that the EUV emission spreads forward whereas  in the higher density quiet-Sun cases the EUV emission brightened at the impact site  and/or  splashed backwards. Probably near the active region the falling material was ionised when still high in the corona. Thus it was trapped and dissipated in the magnetic structures in the corona. The column density of the active-region fragment was relatively high and if falling directly onto the quiet Sun, it should have created a UV kernel.  It may have been ionised either by the increased EUV flux from the active region or shocked, heated and ionised by the splash from an  earlier small impact (see Section~4.3 and Fig~\ref{piglet}). 

Splashes were observed in several of the impacts (1, 4, 11, 12, 13). They are most visible at 171~\AA, implying a temperature slightly under 1~MK. Their plane-of-sky propagation velocity was about 100~\kms\ which is roughly the sound speed in the corona. When impacts fell into a previously created splash, they  dissipated in the corona creating hotter impact emission than in the splash because the falling material was shock heated by the back-flowing high-pressure splash. \citet{Reale14} suggested that the shocked falling plasma may explain the redshifted \CIV\ emission seen in accretion flows. Only the slow, large impact stream produced 1600~\AA\ emission in the falling plasma.  Most the observed fragments have lower column density and larger size than the spherical fragments in the \citet{Reale14} simulations which suggests that they were significantly less dense. This may be the reason that no 1600~\AA\ emission was observed at the head of the falling fragments when they hit a rising splash.

For impacts with velocity greater than about 300~\kms\, 
the biggest influence on the impacts' I$_{UV}$/I$_{EUV}$  was the level of ionisation in the falling fragment since this determined whether the fragment could penetrate to the chromosphere or was trapped and dissipated in the coronal magnetic field. If the fragments reach the chromosphere as predicted by HD simulations \citep{Sacco10, Reale13} then  UV emission which is only weakly absorbed by overlying cool material, should have been bright at all impacts sites. But in situations where the ionisation of the infalling fragment was high, we found only weak UV emission. 
The I$_{UV}$/I$_{EUV}$ decreased with impact velocity which suggests that the ratio is more influenced by impact kinetic energy rather than absorbing column. 
Therefore, we think that if the solar impacts are a good proxy for stellar accretion flows, absorption of EUV and X-ray emission at the impact site is not a likely explanation for the lower accretion rates obtained by analysis of X-ray compared with UV, visible and infrared observations.  
 Rather, we agree with \citet{Bonito14} that a more plausible explanation for the absorption of X-rays is cool  material in the unperturbed accretion stream above the impact site. 
Modelling of both the  UV and EUV emission would give greater understanding of the fragment and impact dynamics.

 
\begin{acknowledgements}
We are indebted to the SDO and STEREO teams for providing the high
resolution data. 
\end{acknowledgements}

\bibliographystyle{aa}

\end{document}